\documentclass[aip,jcp,reprint,groupedaddress]{revtex4-1}

\usepackage{graphicx}
\usepackage{amsmath}
\usepackage{verbatim}
\usepackage{url}
\usepackage[version=3]{mhchem}
\usepackage{chemfig}
\graphicspath{{./pics/}}

\newcommand*{\citen}[1]{%
  \begingroup
    \romannumeral-`\x 
    \setcitestyle{numbers}%
    \cite{#1}%
  \endgroup  
}

\begin{document}


\title{Fabrication of colloidal Laves phases via hard tetramers and hard spheres: bulk phase diagram and sedimentation behaviour}

\author{Guido Avvisati}
\affiliation{Debye Institute for Nanomaterials Science, Utrecht University, Princetonplein 1, 3584CC Utrecht, The Netherlands}

\author{Tonnishtha Dasgupta}
\affiliation{Debye Institute for Nanomaterials Science, Utrecht University, Princetonplein 1, 3584CC Utrecht, The Netherlands}

\author{Marjolein Dijkstra}
\email[]{m.dijkstra@uu.nl}
\affiliation{Debye Institute for Nanomaterials Science, Utrecht University, Princetonplein 1, 3584CC Utrecht, The Netherlands}

\date{\today}

\begin{abstract}
Colloidal photonic crystals display peculiar optical properties which make them particularly suitable for application in different fields. However, the low packing fraction of the targeted structures usually poses a real challenge in the fabrication stage. Here, we propose a novel route to colloidal photonic crystals via a binary mixture of hard tetramers and hard spheres. By combining theory and computer simulations, we calculate the phase diagram as well as the stacking diagram of the mixture, and show that a colloidal analogue of the \ce{MgCu2} Laves phase -- which can serve as a precursor of a photonic bandgap structure -- is a thermodynamically stable phase in a large region of the phase diagram. Our findings show a relatively large coexistence region between the fluid and the Laves phase, which is potentially accessible by experiments. Furthermore, we determine the sedimentation behaviour of the suggested mixture, by identifying several stacking sequences. Our work uncovers a new self-assembly path towards a photonic structure with a band gap in the visible region.
\end{abstract}


\maketitle


\section{\label{sec:intro}Introduction}
It is known that colloidal particles can spontaneously form ordered, periodic phases which are the analogue of crystals in atomic systems. The most prominent example of such a transition, first discovered by computer simulations~\cite{bib:rosenbluth-hs,bib:wood-hs,bib:alder-hs}, and later confirmed by experimental work~\cite{bib:pusey-exp.hs}, is the formation of a Face Centered Cubic (FCC) crystal from a fluid of colloidal particles which behave as Hard Spheres (HS). 

The study of crystalline phases on colloidal length and time scales is important not only at a fundamental level, where it allows for insights into e.g., phase transitions and crystallisation kinetics~\cite{bib:anderson-rev.colloids,bib:weitz-rev.colloids,bib:li-rev.colloids}, but also for potential applications. In particular, it is possible to fabricate photonic crystals (PCs) from colloidal particles. By PCs we mean structures with a periodically varying dielectric constant that display a complete photonic band gap. Due to to the intrinsic size of the employed building blocks, colloidal photonic crystals display a band gap in the visible range of frequencies. These structures act for photons in the same way as semiconductors do for electrons, hence opening up a way to control light propagation. The application area of such materials is quite broad, ranging from optical fibers, displays and switches to (bio-)sensing and bio-medical engineering, and finally to energy storage and security~\cite{bib:stein-rev.pc,bib:ozin-rev.pc,bib:cong-rev.pc,bib:xu-rev.pc,bib:zhang-rev.pc,bib:lopez-rev.pc.1,bib:lopez-rev.pc.2,bib:ge-rev.pc,bib:sawada-colloids.photonics,bib:cullen-blast.sensor,bib:lee-pc.k.security,bib:burgess-pc.wink.security}. Therefore, a significant amount of research in the colloid science community deals with the design and fabrication of such photonic crystals. 

Since the early work on PCs~\cite{bib:bykov-theory.pbg,bib:yablo-theory.pbg1,bib:yablo-theory.pbg2,bib:yablo-theory.pbg3}, different particle arrangements were explored as candidates~\cite{bib:ho-theory.bg.dc,bib:ho-theory.woodpile,bib:sozuer-theory.woodpile,bib:sozuer-theory.bg.bcc,bib:sozuer-theory.bg.fcc,bib:joannopoulos-theory.bg.rods}, and some of them -- most notably the so called ``inverse opals'' -- were also fabricated in the lab~\cite{bib:lin-woodpile,bib:fleming-woodpile,bib:wijnhoven-inverse.opal,bib:stein-inverse.pcs,bib:vlasov-inverse.opal,bib:qi-rod.stack}. To date, the most suitable structures to make PCs remain the Diamond Crystal (DC) and the Pyrochlore structure, in which the colloids are located on the lattice positions of the respective crystal structures~\cite{bib:maldovan-diamond.crystals,bib:vermolen-diamond.pyro}. However, despite the efforts, the fabrication of such open (non close-packed) structures at the colloidal scales has not been achieved yet, and it is a long-standing research focus in the nanomaterials and colloid science community.

Nevertheless, new perspectives on the subject arise because the recent advances in the colloidal synthesis allow for more and more exotic building blocks to be used in the colloidal self-assembly arena. Clusters of spheres with well-defined shapes, such as dimers, trimers and tetramers, have become available, together with the intriguing possibility of employing them to self-assemble PCs~\cite{bib:pine-clusters.1,bib:pine-clusters.3,bib:kraft-colloidal.molecules,bib:summers-tetramers.for.pc,bib:yang-rev.multimers,bib:yin-pc.from.clusters,bib:wang-review.microflu,bib:sacanna-review.colloids}. These colloidal clusters can be produced in several ways. One method takes advantage of the drying forces in an evaporating emultion droplet to drive the confined colloidal particles to a specific geometry~\cite{bib:pine-clusters.1,bib:pine-clusters.3,bib:yang-rev.multimers}. A different class of fabrication procedures relies instead on microfluidics setups, with of without the use of lithographically patterned surfaces.~\cite{bib:xia-cluster.microflu,bib:yin-cluster.microflu,bib:weitz-cluster.microflu,bib:wang-review.microflu,bib:ravaine-review.coll.molecules}.

In addition, on the theoretical side, two new ideas were put forward to possibly facilitate the fabrication of PCs, and we shall briefly discuss them in the following. One study showed that a structure composed of tetrahedral clusters of spheres (``tetrastack'') displays a photonic band in the optical region~\cite{bib:joannopoulos-tetrastack}. However, while they employ a complex building block, it is not clear how the suggested structure can be realised experimentally. Another study suggested that, by using a binary mixture of colloidal particles with different sizes, it is possible to assemble an \ce{MgCu2} Laves phase. This is appealing because the \ce{MgCu2} consists of a DC of large spheres and a Pyrochlore lattice of small spheres, and both substructures display a photonic bandgap~\cite{bib:dijkstra-laves.short}. In this case, the authors addressed the problem posed by the open structure by using a binary mixtures of spheres. Nevertheless, issues arise when one considers that three phases can actually be assembled from a binary hard-sphere mixture, namely the \ce{MgCu2}, the \ce{MgNi2}, and the \ce{MgZn2}. It is also important to note that the latter is the thermodynamically stable phase, and unfortunately not the aimed \ce{MgCu2} phase~\cite{bib:dijkstra-laves.long}. Furthermore, the three aforementioned Laves phases are nearly degenerate as they have very similar free energies, hence the self-assembly of the mixture results in glassy states, unless the assembly is directed, \emph{e.g.}, by using templated walls~\cite{bib:dijkstra-laves.short}.

In this work, we combine Monte Carlo (MC) computer simulations and theory to study the phase behaviour of a binary mixture of large hard spheres and rigid, hard tetrahedral clusters of small hard spheres (hereafter denoted as tetramers) with a fixed size ratio. For this mixture, we compute both the bulk phase diagram and the sedimentation behaviour. In particular, using free-energy calculations, we address the stability of the \ce{MgCu2} Laves phase that can result from the self-assembly of the mixture. In this way, we retain the best of both approaches previously introduced, while also circumventing some of the other problems.

For instance, employing a binary mixture mitigates the problem of the low-coordinated open target structures of the diamond and pyrochlore phase, whereas using tetramers as one of the building blocks alleviates the lattice degeneracy problem as \ce{MgNi2} phase cannot be self-assembled from tetramers and spheres, and moreover using tetramers also removes the metastability problem as the \ce{MgCu2} phase is more stable than the \ce{MgZn2} phase in mixtures of tetramers and spheres.  Hence, the particular choice of colloidal building blocks intrinsically pre-selects the desired structure, and thus the \ce{MgCu2} Laves phase is obtained by design. Furthermore, by using the bulk phase diagram and the local density approximation, we theoretically calculate the stacking diagram of the mixture, which predicts the stacking sequences of different phases that could be observed in sedimentation experiments on the same mixture. 

We stress that such a model mixture is well within experimental reach, even though no studies on it have been performed yet, to the best of our knowledge. This is somewhat surprising as hard-core systems are usually much easier to control than systems with attractive interactions, which often requires substantial fine-tuning of the range, strength, and directionality of the interactions.

The paper is organised as follows. We introduce the model and discuss the employed methods in Sec.~\ref{sec:model}. In Sec.~\ref{sec:res1} we present the results on the phase behaviour of the binary mixture of spheres and tetramers, while in Sec.~\ref{sec:res2} we discuss the sedimentation behaviour. In Sec.~\ref{sec:summary} we sum up our findings, and outline future research directions.

\section{\label{sec:model}Model and Methods}
\subsection{\label{ssec:sims} Monte carlo simulations}
\begin{figure}[htb]
  \includegraphics[scale=0.14]{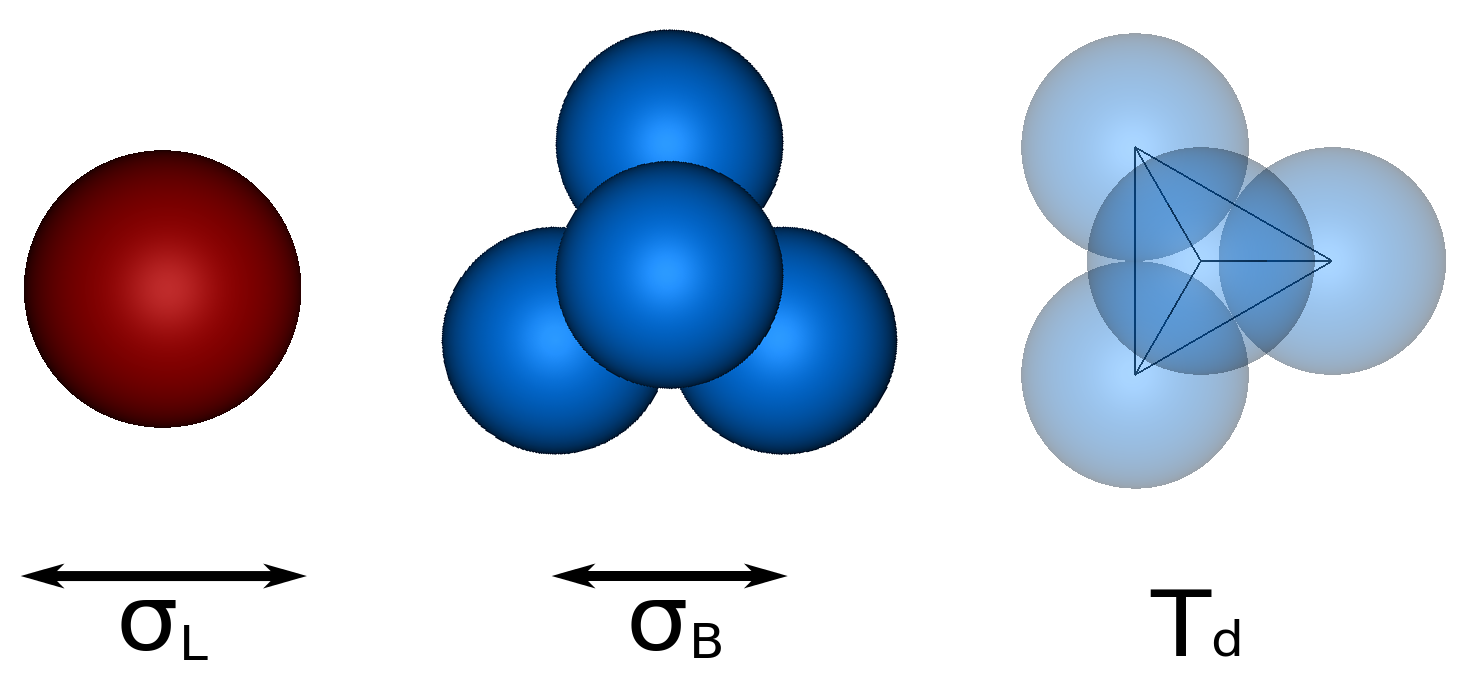}
  \caption{{\small Building blocks of the investigated binary mixture. (left) Hard spheres with diameter $\sigma_L$. (center) Hard tetrahedral tetramers of bead size $\sigma_B$. Note that the beads are tangential to one another. The size ratio $q=\sigma_B/\sigma_L$ is fixed to 0.82. (right) Faceted model of tetrahedron, with symmetry group $T_d$, connecting the centers of the beads. }}
  \label{fig:mix.model}
\end{figure}

We consider a binary mixture of $N_s$ hard spheres and $N_t$ hard tetramers with composition $x=N_s/N$, where $N=N_s+N_t$. The spheres have diameter $\sigma_L$. Each tetramer consists of four touching spherical beads of diameter $\sigma_B$ arranged in a tetrahedral fashion. We assume the tetramers to behave like a rigid body, \emph{i.e.}, fluctuations in the geometrical arrangement of the spheres are neglected. The size ratio between a bead in a tetramer and a sphere is labelled as $q=\sigma_B/\sigma_L$. Since the \ce{MgCu2} Laves phase of an ordinary binary hard-sphere mixture achieves its highest packing fraction for $q=\sqrt{2/3}\sim 0.82$~\cite{bib:liaw-laves.size.ratio,bib:edwards-laves.size.ratio,bib:dijkstra-laves.short,bib:dijkstra-laves.long}, we employ this value in our work. All interactions are assumed to be HS-like, meaning that the objects do not interpenetrate each other. Thus, spheres cannot approach each other closer than $\sigma_L$, beads belonging to different tetramers cannot approach each other closer than $\sigma_B$, spheres and tetramer beads cannot approach closer than $\sigma_{LB}=(\sigma_L+\sigma_B)/2$. A model of the different building blocks employed in this work is shown in Fig.~\ref{fig:mix.model}.

In order to map out the phase diagram of the system, we combine Monte Carlo (MC) simulations in the isobaric-isothermal ensemble and free-energy calculations. Hence, the relevant thermodynamic quantities are $N_s,N_t,P,T$.  The pressure $P$ is measured in reduced units as $\beta P\sigma_L^3$ with $\beta = 1/k_{\mathrm{B}}T$, $k_{\mathrm{B}}$ Boltzmann's constant, and $T$ the system temperature. The packing fraction is defined as $\eta=\gamma\rho$, where $\rho=N/V$ is the number density, $V$ the volume of the simulation box, and $\gamma=\pi\sigma_L^3[x+4q^3(1-x)]/6$. To evolve the system, we use displacement moves for spheres and tetramers, rotational moves for tetramers, and volume moves. For each move, we set an acceptance rate of 30\%. An MC step (MCS) is defined as $N$ attempted translations or rotations, and one volume move of the simulation box. The length of the simulations in the isobaric-isothermal ensemble is at least $5\times 10^6$ MCS, while the free-energy calculations run for at least $2\times 10^6$ MCS. For the case of non-cubic crystal structures, we also employ $NPT$ simulations where the box lattice vectors are free to fluctuate, in order to remove any additional stress from the crystal phase~\cite{bib:parrinello-floppy.box,bib:filion-floppy.box}. For each composition of large spheres $x$, the equation of state (EOS) is computed by means of compression and expansion runs. For the compression runs, the starting configuration is a disordered fluid of $N_s=xN$ spheres and $N_t=(1-x)N$ tetramers. For the expansion runs, crystalline structures of selected composition provide the initial configuration as explained in the following.

\subsection{\label{ssec:crystals}Crystalline structures}
For a binary hard-sphere mixture, previous studies have shown that, at the chosen size ratio $q=\sigma_B/\sigma_L=0.82$, the stable crystal structures are the pure FCC crystals of large and of small spheres, and the \ce{MgNi2}, \ce{MgCu2}, and \ce{MgZn2} Laves phases~\cite{bib:dijkstra-laves.short,bib:dijkstra-laves.long}, where the \ce{MgZn2} phase has a slightly lower free energy than the other two, and the \ce{MgCu2} Laves phase can be stabilised by wall templating~\cite{bib:dijkstra-laves.short}. In the case of a mixture of tetramers and spheres, we employ the same packing arrangements as those in Ref.~\citen{bib:dijkstra-laves.long}, but we replace four small spheres by a tetramer. This procedure yields structures which are made from the investigated building blocks (spheres and tetramers), but are arranged similarly to the respective literature cases. In particular, the FCC of small spheres at $x=0$ becomes a simple cubic crystal lattice of tetramers. Furthermore, it is important to note that the third kind of Laves phase -- the \ce{MgNi2} crystal -- cannot be reproduced by a combination of tetramers and spheres, hence it falls already out of the picture when considering candidate crystal structures.
\begin{figure}[htb]
  \includegraphics[width=0.45\textwidth]{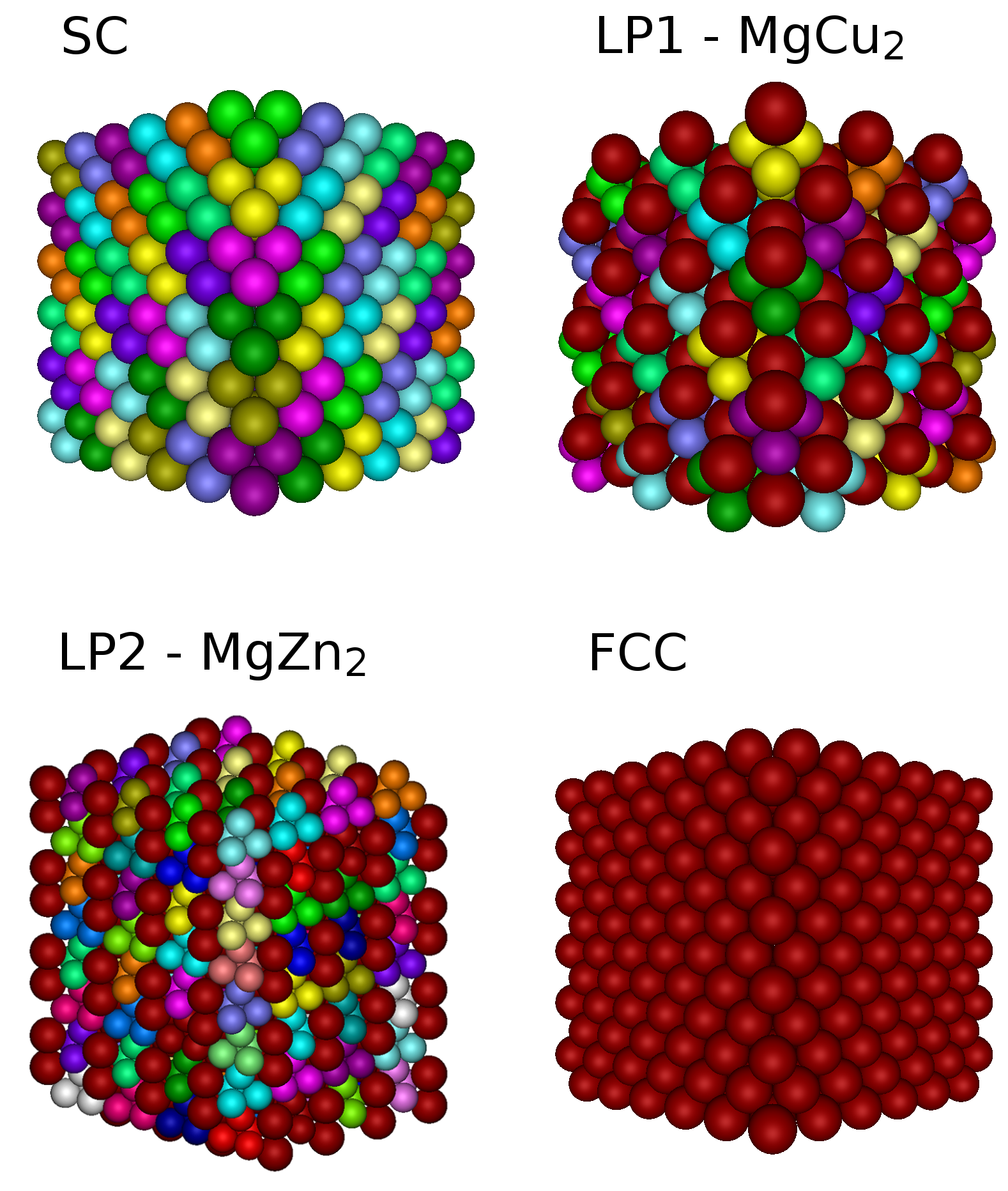}
  \caption{{\small Crystal structures considered in this work. (top left) The Simple Cubic crystal of tetramers (SC) at composition $x=0$. (top right) The binary \ce{MgCu2} Laves phase (LP1) at composition $x=2/3$. (bottom left) The binary \ce{MgZn2} Laves phase (LP2) at composition $x=2/3$. (bottom right) The Face Centered Cubic crystals of large spheres (FCC) at composition $x=1$. The color code identifies different tetramers and separates tetramers from spheres.}}
  \label{fig:xtals.intro}
\end{figure}

\noindent
Summing up, for different compositions of large spheres $x$ we have:
\begin{itemize}
\item[\textbf{SC\phantom{a}}] simple cubic lattice of tetramers with specified orientation at composition $x=0$.
\item[\textbf{LP1}] a mixed structure of tetramers and spheres, which packs the same way as an ordinary \ce{MgCu2} lattice, at composition $x=2/3$.
\item[\textbf{LP2}] the analogue of the \ce{MgZn2} crystal, but made out of tetramers and spheres, also at composition $x=2/3$. Note that this structure has a non-cubic unit cell.
\item[\textbf{FCC}] the thermodynamic stable structure for hard spheres, at composition $x=1$.
\end{itemize}

\noindent
In the SC, LP1 and LP2 phases, respectively, all the tetramers have the same orientation, which is calculated by a rigid transformation of the bead positions in the reference frame to the bead positions in the crystal at hand. We note that other arrangements are, in principle, possible for the SC phase, with respect to both the positions and the orientations of the tetramers, nevertheless the positions of the beads of the tetramers must always be compatible with an FCC packing. Moreover, the degeneracy of the SC phase, if present at all, is expected to be small~\cite{bib:kowalik-tetramer.solids}, hence we neglect it in our calculations. 

\subsection{\label{ssec:energymc}Free-energy calculations}
The bulk phase diagram is determined by using the common tangent construction in the Gibbs free energy $g$ -- composition $x$ representation. We remind the reader that the dimensionless Gibbs free energy per particle is defined as $g=\beta G/N = f + Z$, where $f = \beta F/N$ is the dimensionless Helmholtz free energy per particle and $Z=\beta P/\rho=\gamma\beta P/\eta$ is the compressibility factor. 

Thus, in order to compute the Gibbs free energy $g$, one must first calculate $f$, and thermodynamic integration is the method of choice for this task~\cite{bib:frenkel-ums}. Starting from a reference point, $f$ is obtained by integrating the EOS to the point of interest, assuming no phase transition is crossed along the integration path
\begin{equation}
f(\eta)=f(\eta_0)+\frac{\gamma}{k_{\mathrm{B}}T}\int_{\eta_0}^{\eta}d\eta'\frac{P(\eta')}{\eta'^2}
\label{eq:therm.int}
\end{equation}

The main problem is now shifted to the computation of $f$ at the reference point. For the fluid phase we choose this point to be an ideal gas mixture. For the crystal phases we use the Frenkel-Ladd method extended to account for the anisotropic particle shape~\cite{bib:frenkel-ums,bib:frenkel-einstein.xtal,bib:dijkstra-hard.dumbbells,bib:vega-review.free.ene}. In this method, one connects an Einstein crystal, where particles are tied to their ideal lattice positions and orientations by harmonic springs, to the system of interest by slowly removing the harmonic springs. More details can be found in Ref.~\citen{bib:vega-review.free.ene} and references therein. The Helmholtz free-energy per particle $f$ of a crystal reads~\cite{bib:dijkstra-hard.dumbbells,bib:vega-review.free.ene}:
  \begin{equation}
    f(\eta_0)=f_{\textrm{Einst}}(\lambda_{\textrm{max}})-\frac{1}{N}\int_0^{\lambda_{\textrm{max}}}d\lambda'\left\langle\frac{\partial\beta U_{\textrm{Einst}}(\lambda')}{\partial\lambda'}\right\rangle_{\textrm{NVT}}
    \label{eq:einst.xtal}
  \end{equation}
where $f_{\textrm{Einst}}$, which stands for the free energy per particle of an ideal Einstein crystal, is given by:
\begin{widetext}
  \begin{equation}
    f_{\textrm{Einst}}(\lambda_{\textrm{max}})=-\frac{3(N-1)}{2N}\ln\left(\frac{\pi}{\lambda_{\textrm{max}}}\right)+\ln\left(\frac{\Lambda_t^3\Lambda_r}{\sigma_L^3}\right)+\frac{1}{N}\log\left(\frac{\sigma_L^3}{VN^{1/2}}\right)+(1-x)f_{\textrm{or}}(\lambda_{\textrm{max}})
    \label{eq:einst.ideal}
  \end{equation}
\end{widetext}
In Eq.~\ref{eq:einst.xtal}, the function $U_{\textrm{Einst}}(\lambda)$ denotes the harmonic potential that couples the particles positions and orientations to the corresponding Einstein lattice values and reads:
\begin{equation}
  \begin{split}
    \beta U_{\textrm{Einst}}(\lambda)\;=\;&\lambda\sum_{i=1}^N\left(\boldsymbol{r}_i-\boldsymbol{r}_{i,0}\right)^2/\sigma_L^2\;+\\
    &\lambda\sum_{i=1}^{N_t}\left(\sin^2\psi_{ia}+\sin^2\psi_{ib}\right)
  \end{split}
  \label{eq:einst.ene}
\end{equation}
where $\left(\boldsymbol{r}_i-\boldsymbol{r}_{i,0}\right)$ represents the displacement of particle $i$ from its position in the ideal Einstein crystal, and where the angles $\psi_{ia}$ and $\psi_{ib}$ are the minimum angles formed by the vector pointing to any of the beads in the tetramer and the rest position of two arbitrarily chosen beads $a$ and $b$, respectively. Note that all the spheres and tetramers are connected with springs to their respective lattice positions in the Einstein crystal, whereas an aligning potential is acting only on the tetramers. The term $f_{\textrm{or}}(\lambda_{\textrm{max}})$ in Eq.~\ref{eq:einst.ideal} takes into account the orientational free energy of the ideal Einstein crystal and reads:
\begin{equation}
  \begin{split}
    f_{\textrm{or}}(\lambda_{\textrm{max}})=&-\ln\left\{\frac{1}{8\pi^2}\int d\phi d\theta d\chi\sin(\theta)\;\times \right.\\
    &\left. \exp\left[-\lambda_{\textrm{max}}\left(\sin^2\psi_{ia}+\sin^2\psi_{ib}\right)\right]\right\}
  \end{split}
  \label{eq:einst.ideal.or}
\end{equation}
where $\phi$, $\theta$ and $\chi$ are the Euler angles. This integral depends only on the maximum value chosen for the coupling constant $\lambda$ and, of course, on the form of the Hamiltonian chosen for the orientational springs. In simple cases, it can be evaluated exactly or in an approximated analytic form. However, when the orientational Hamiltonian is more complex as in the current case, it must be calculated numerically, e.g. via MC integration. 

Once the Helmholtz free energy is known, the Gibbs free energy per particle for fixed composition and varying pressure is calculated as
\begin{equation}
  g(P,x) = f(\eta_0,x) + \gamma\int_{\eta_0}^{\eta}d\eta'\frac{\beta P(\eta',x)}{\eta'^2} + Z(P,x)
  \label{eq:gibbs.free.ene}
\end{equation}

\begin{figure}[htb]
  \includegraphics[scale=0.33]{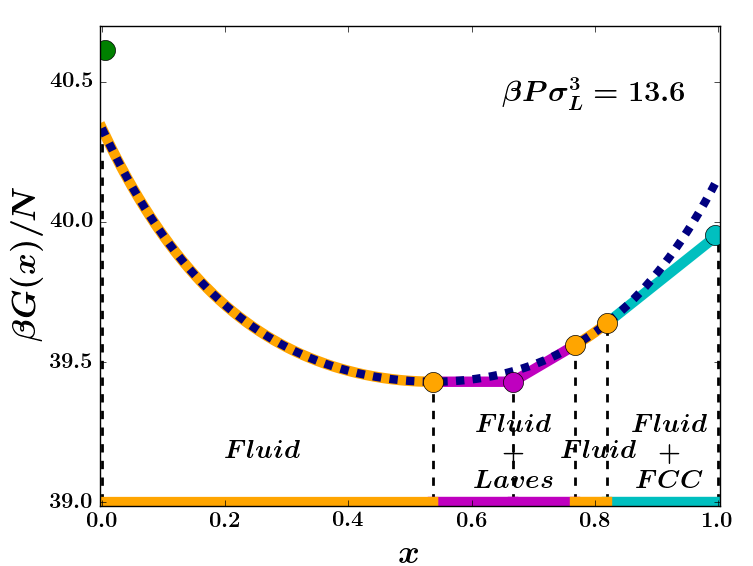}
  \caption{{\small Gibbs free energy per particle $g=\beta G(P,x)/N$ as function of composition $x=N_s/N$ for a fixed pressure $\beta P\sigma_L^3=13.6$. The green, magenta and cyan dots represent the SC phase of pure tetramers (at $x=0$), the LP1 crystal (at $x=2/3$), and the FCC of pure large spheres (at $x=1$), respectively. The blue dashed line shows the Gibbs free energy $g(P,x)$ of the fluid as function of composition $x$. The orange dots represent the coexistence points between the fluid and the LP1 crystal (2 points), and between the fluid and the FCC crystal of pure large spheres, as calculated by the common tangent construction. The thick lines show the path of minimal Gibbs free energy.}}
  \label{fig:gfe.sample}
\end{figure}

With the outlined procedure, we calculate the Gibbs free energy $g(P,x)$ for the fluid phase at different compositions with a grid spacing of $0.1$, as well as the Gibbs free energy $g(P,x)$ for the solid phases. We then use the common tangent construction in the $(g,x)$-plane to draw the phase diagram. A representative calculation of $g(P,x)$ is given in Fig.~\ref{fig:gfe.sample}, where we also show the results of the common tangent construction. By collecting the information about $g(P,x)$ at several pressure values, we eventually map out the phase diagram of the binary mixture in the pressure $\beta P\sigma_L^3$ -- sphere composition $x$ representation.

\subsection{\label{ssec:stackmeth}Stacking sequences and stacking diagram}
Once ascertained the bulk thermodynamics, we also study the system sedimentation behaviour. To this end, we theoretically construct a stacking diagram which is the set of all possible sequences of phases stacked in a sedimentation column, following the method recently presented in Ref.~\citen{bib:delasheras-stacking.diagrams}. The theory behind the construction of a stacking diagram is based on chemical potentials, hence the bulk phase diagram in the $P-x$ representation must first be converted to the plane of chemical potential of the spheres (L) and tetramers (T), respectively. In the following, we assume that such a conversion has been done and only discuss in terms of chemical potentials of the two species. 

Once gravity is switched on, there is an increasing concentration profile along the column in the direction of gravity $z$. We now define a $z$-dependent chemical potential which varies with concentration along the sedimentation column 
\begin{equation}
\psi_i(z) = \mu_i^{0} - m_igz
\label{SD1}
\end{equation}
where $\psi_i(z)$ is the chemical potential of species $i=\{L,T\}$ at a height $z$ of the column , $\mu_i^{0}$ is its chemical potential in the absence of gravity, and $m_i$ its buoyant mass. Rearranging Eq.~\ref{SD1} and eliminating the $z$-dependence, we obtain a linear relation between the chemical potential of the spheres $\psi_L(z)$ and the chemical potential of the tetramers $\psi_T(z)$
\begin{align}
  \psi_L(\psi_T) &= a + s \psi_T \label{SD2}\\
  s &= m_L/m_T \label{SD3}\\
  a &= \mu_L^{0} - s\,\mu_T^{0}\label{SD4}
\end{align} 
where $s$ is the gravitational variable and $a$ is the composition variable. 

Assuming the local density approximation (LDA) is valid, we can set the local chemical potential $\psi_i(z)$ of species $i$ equal to the chemical potential $\mu_i$ of an equilibrium bulk state, i.e
  \begin{equation}
   \psi_i(z)=\mu_i
   \label{SD5}
  \end{equation} 
so that the correlation between $s$ and $a$ appears as a straight line (Eq.~\ref{SD2}) on the plane of chemical potentials $\mu_T-\mu_L$. This straight line is called a ``sedimentation path" and the set of all such lines constitutes a stacking diagram. The point at which a sedimentation path crosses a bulk binodal represents a phase transition. Therefore, each path yields a specific stacking sequence of phases in the corresponding stacking diagram. 

\section{\label{sec:res1}Bulk phase behaviour}
In this section, we present and discuss our results for the bulk phase diagram of the binary mixture of spheres and tetramers, including a representation of the phase diagram more suitable to experiments.

\subsection{\label{ssec:bulkeos}Equations of state}
\begin{figure}[htb]
  \centering
  \includegraphics[width=0.45\textwidth]{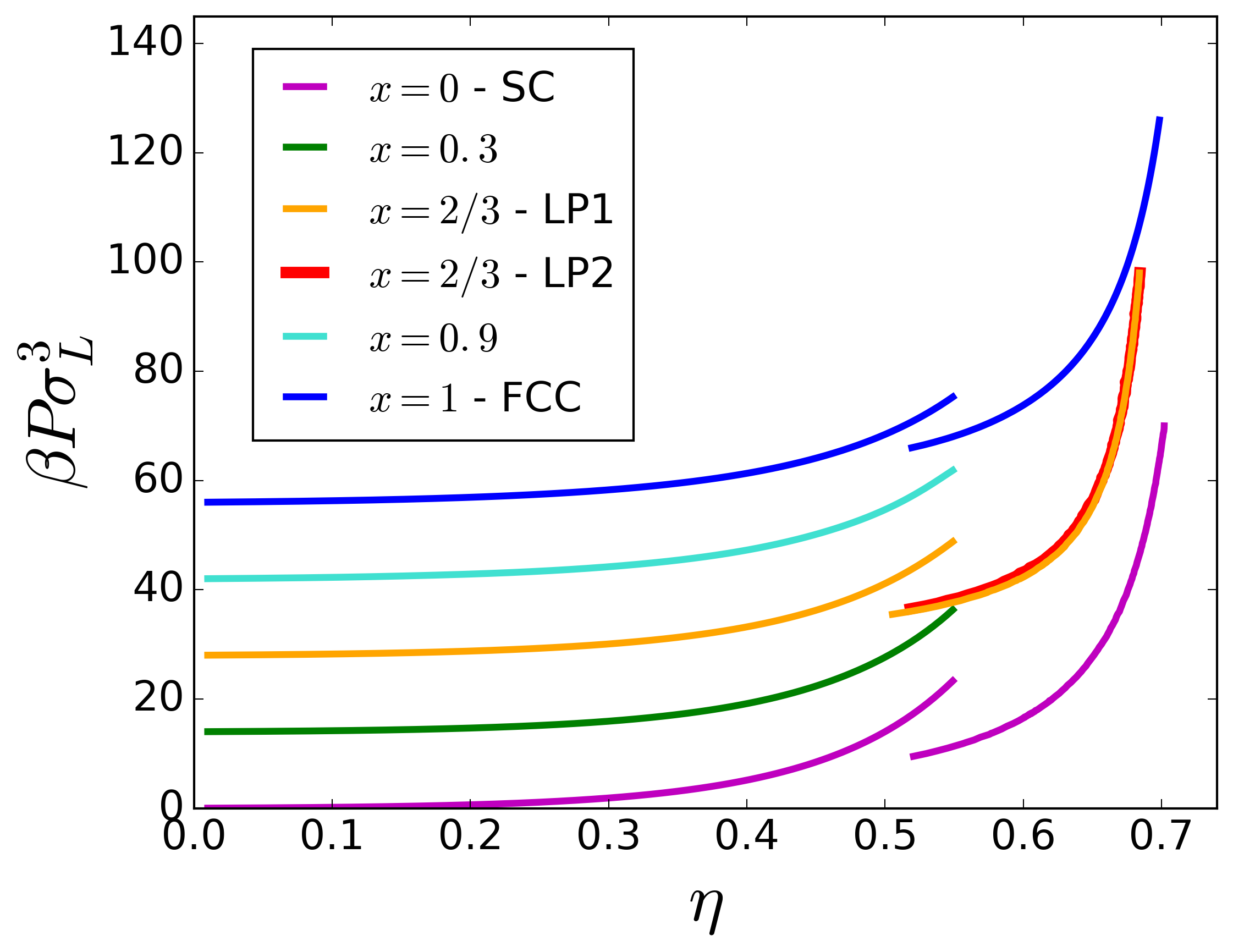}
  \caption{{\small EOS of the mixture of large hard spheres and hard tetramers at different compositions $x$. The solid branches correspond to expansion runs of the studied solid structures, namely SC at $x=0$, LP1 and LP2 at $x=2/3$, and FCC at $x=1$. Note that the EOSs of LP1 and LP2 coincide for high pressures, but differ at lower pressures. For visualisation purposes, the curves have been shifted with respect to each other in the $y$ direction by an amount $\Delta y=14$.}}
  \label{fig:eos.system}
\end{figure}

The equations of state (EOS) of both the fluid phase at different compositions $x$ and of the crystalline structures considered are a key ingredient of the calculation of the phase diagram, as we see from Eq.~\ref{eq:gibbs.free.ene}. For the fluid phase, we calculated the EOSs at composition intervals of $0.1$, whereas for the crystal phases the compositions are fixed. In Fig.~\ref{fig:eos.system} we show the EOSs of the different crystal structures investigated, as well as the EOSs of the fluid mixture at different compositions $x$. 
We subsequently fit the simulation results to 
\begin{equation}
  \frac{\gamma\beta P}{\eta} = 1 + \sum_{i=1}^n a_i \eta^i
  \label{eq:compflu}
\end{equation}
for the fluid phase, and
\begin{equation}
  \frac{\gamma\beta P}{\eta^2} = \sum_{i=0}^m b_i \eta^i
  \label{eq:compxtl}
\end{equation}
for the crystal phases. The typical value of $n$ is $12$, while $m=3$ for all cases. The fitting procedure allows us to easily perform the thermodynamic integration in Eq.~\ref{eq:gibbs.free.ene}.

\subsection{\label{ssec:bulkpd}Stabily of LP1--\ce{MgCu2} and phase diagrams}
\begin{figure}[htb]
  \centering
  \includegraphics[width=0.45\textwidth]{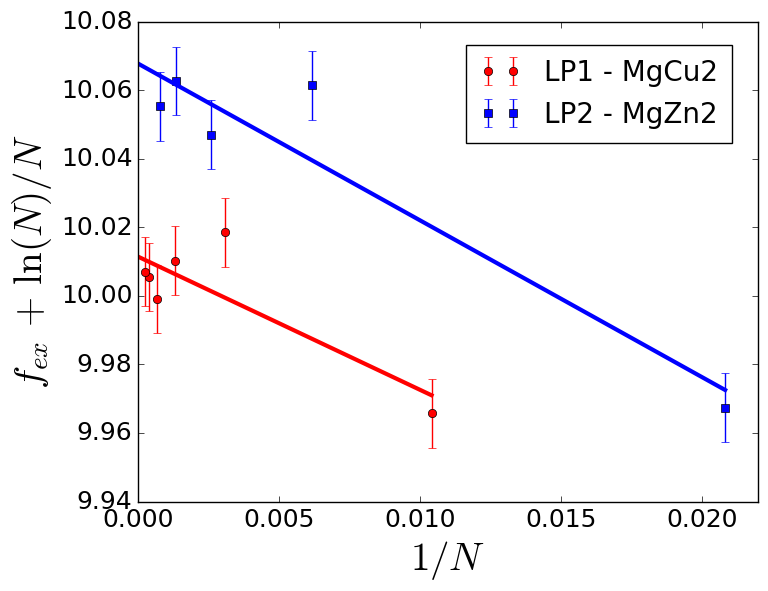}
  \caption{{\small Finite-size scaling of the excess Helmholtz free energy $f_{ex}\,+\,\ln(N)/N$ vs $1/N$ at diamter ratio $q=0.82$ and packing fraction $\eta=0.6$ for the LP1--\ce{MgCu2} and LP2--\ce{MgZn2} Laves structures of hard tetramers and hard spheres. The lines are linear fits to the data points. the LP1 crystal is always significantly more stable than the LP2 structure, the free energy difference in the thermodynamics limit being $6\times 10^{-2}$ per particle.}}
  \label{fig:laves.fss}
\end{figure}

Previous work on binary hard-sphere mixtures has shown that, unless wall templating is used, the \ce{MgZn2} Laves phase is more stable than the \ce{MgCu2} Laves phase~\cite{bib:dijkstra-laves.short,bib:dijkstra-laves.long}. Unfortunately, the \ce{MgCu2} structure is the only Laves phase whose sublattices display a complete photonic band gap~\cite{bib:maldovan-diamond.crystals,bib:vermolen-diamond.pyro}. Hence, the first issue for us to investigate is the thermodynamic stability of LP1--\ce{MgCu2} compared to LP2--\ce{MgZn2}. We addressed this by performing free-energy calculations at a fixed packing fraction of $\eta=0.60$ for different total number of particles $N$. By plotting the excess free energy per particle including finite-size corrections versus $1/N$ for both structures, we can extrapolate to the thermodynamic limit $N\rightarrow\infty$ by looking at the intercept of the two lines~\ref{bib:frenkel-ums,bib:polson-fss,bib:vega-review.free.ene}. This is displayed in Fig.~\ref{fig:laves.fss}, where it becomes clear that the LP1--\ce{MgCu2} structure of hard tetramers and hard spheres is more stable than the LP2--\ce{MgZn2} structure in the thermodynamic limit. The LP1 structure has a bulk excess free energy per particle of 10.01(1)$k_{\textrm{B}}T$, while the LP2 crystal has an excess free energy per particle of 10.07(1)$k_{\textrm{B}}T$, the difference being $6\times 10^{-2}k_{\textrm{B}}T$ per particle. Incidentally, we note that this free-energy difference is not at all small, being hundreds of times larger than the free-energy difference per particle between an FCC and an HCP of hard spheres. Thus, by employing a mixture of hard tetramers and large hard spheres, the \ce{MgCu2} structure -- the precursor of colloidal photonic crystals -- is stabilised in bulk. In view of this result, we will refer to LP1 more generically as ``Laves phase'' in the following.

\begin{figure}[htb]
  \centering
  \includegraphics[width=0.5\textwidth]{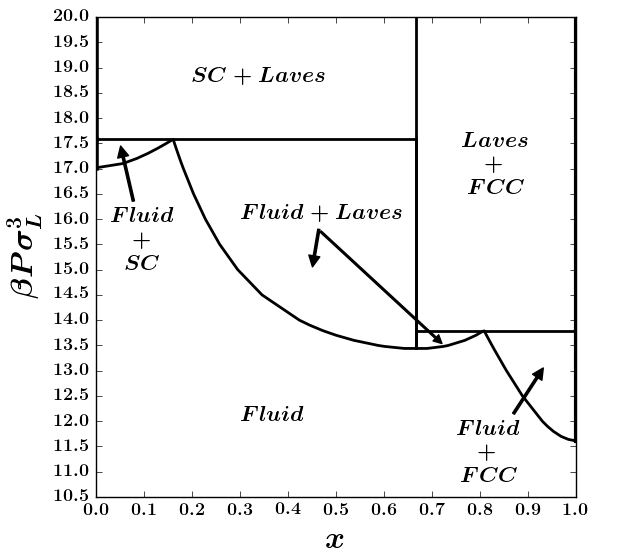}
  \caption{{\small Phase diagram of a binary mixture of hard spheres and hard tetramers in the pressure $\beta P\sigma_L^3$--composition $x$ representation. The composition $x=N_S/N$ refers to the spheres. Two triple points (Fluid+SC+Laves, Fluid+Laves+FCC) are found, together with a relatively large phase coexistence region between the fluid and the Laves phase.}}
  \label{fig:px.diagram}
\end{figure}

To draw the phase diagram in the pressure $\beta P\sigma_L^3$--composition $x$ representation, we apply common tangent constructions to the Gibbs free-energy curves $g(P,x)$ at different pressures, in order to determine the composition and densities of the coexisting phases. The results are summarised in Fig.~\ref{fig:px.diagram}. 

For pressures $\beta P\sigma_L^3\leq 11.5$, we find that the fluid is the only stable phase. Increasing the pressure results in different coexistence regions, between the fluid and the three crystal structures investigated, and between the different crystal structures at even higher pressures. 

For $11.5 \leq \beta P\sigma_L^3\leq 13.9$ and compositions $x>0.81$ we find coexistence between the FCC crystal of large spheres and the fluid phase, while for $17.0 \leq \beta P\sigma_L^3\leq 17.6$ and compositions $x<0.17$ we find a coexistence between the simple cubic crystal of tetramers and the fluid phase. 

Interestingly, at intermediate pressures and compositions we observe two distinct phase coexistence regions between the Laves phase and the fluid phase with either a composition smaller or larger than that of the Laves phase, \emph{i.e.}, $x\leq 2/3$ and $x\geq 2/3$. Moving towards high pressures we find solid-solid coexistence between the simple cubic phase of pure tetramers and the Laves phase, and between the Laves phase and the pure FCC phase of large spheres, the former starting at somewhat higher pressures than the latter ($\beta P\sigma_L^3 > 17.6$ vs $\beta P\sigma_L^3 > 13.9$). 

For very high pressures, we expect, due to packing considerations, only a single coexistence region between the simple cubic phase of tetramers and the FCC crystals of large spheres, i.e. we expect to find another triple point where the SC, Laves, and FCC phases are in coexistence with each other. However, we were unable to detect the crossover, even by simulating at pressures as high as $\beta P\sigma_L^3 = 70.0$. Thus, we can only set a lower limit on this specific crystal-crystal phase coexistence region. 

\begin{figure}[htb]
  \includegraphics[scale=0.225]{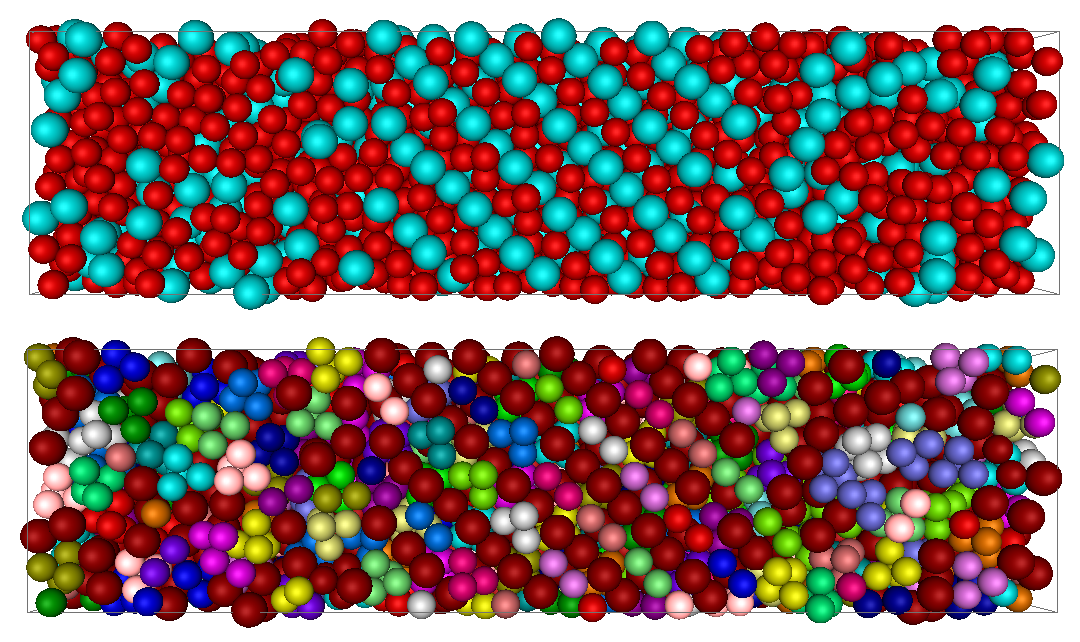}
  \caption{{\small Representative final configuration from direct coexistence simulations displaying coexistence between the fluid phase and the Laves crystal of hard tetramers and hard spheres. (top) Overall composition $x=0.6$ and pressure $\beta P \sigma_L^3=15.0$. (bottom) Same as top panel, but with color coding as to highlight the different tetramers.}}
  \label{fig:coex.pic}
\end{figure}

The relatively large two-phase coexistence region between the fluid phase and the Laves phase is the most remarkable feature of the presented phase diagram, signalling an extended and easily accessible parameter range to obtain the targeted \ce{MgCu2} Laves phase in simulations as well as in experiments. We checked this result by additionally performing direct coexistence simulations at overall compositions $x=0.5$ and $x=0.6$ and pressure $\beta P \sigma_L^3=15.0$. In Fig.~\ref{fig:coex.pic} we present snapshots of the final configurations as obtained from the simulations, which confirm the coexistence between the fluid phase and the Laves phase of tetramers and spheres.

\begin{figure}[htb]
  \includegraphics[scale=0.35]{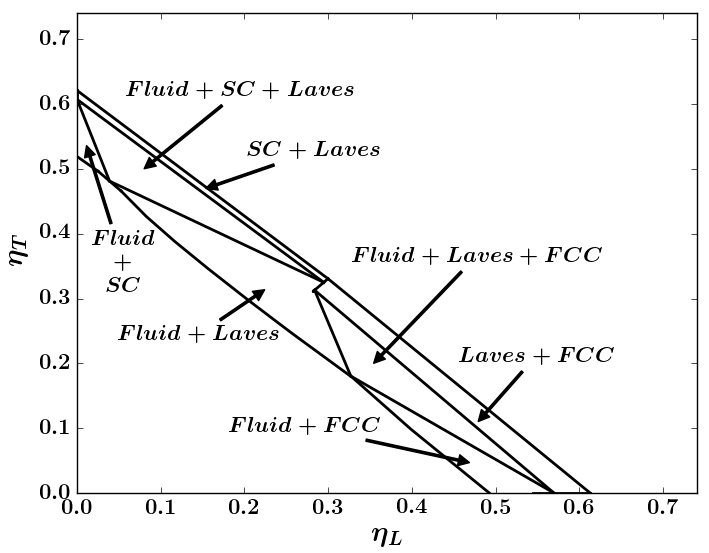}
  \caption{{\small Phase diagram of the investigated binary mixture in the packing fraction of tetramers $\eta_T$ -- packing fraction of large spheres $\eta_L$ representation.}}
  \label{fig:eta2.plot}
\end{figure}
Despite the progress in the fabrication of colloidal building blocks, we are unaware, to the best of our knowledge, of experimental realisations of the proposed binary mixture. In order to facilitate the comparison with experimental results we additionally convert the phase diagram to the packing fraction of tetramers $\eta_T$ -- packing fraction of spheres $\eta_S$ representation, the result being shown in Fig.~\ref{fig:eta2.plot}. The triple points we found in Fig.~\ref{fig:px.diagram} -- Fluid + SC + Laves, Fluid + Laves + FCC -- transform to triangular areas in this representation. In between the triple points we find the coexistence region between fluid phase and Laves structure, which could be probed experimentally. Note that the triple point SC + Laves + FCC is outside the scanned pressure range and for this reason does not appear in Fig.~\ref{fig:eta2.plot}.
Finally, we also calculate the phase diagram in the chemical potential of the spheres $\mu_L$ -- chemical potential of the tetramer $\mu_T$ representation. While this diagram is not suitable for comparison with experiments, it is instead crucial in order to theoretically address the role of gravity on the presented bulk results, as accomplished in the next section.

\section{\label{sec:res2}Sedimentation behaviour and stacking diagram}
\begin{figure}[htb]
  \centering
  \includegraphics[width=0.45\textwidth]{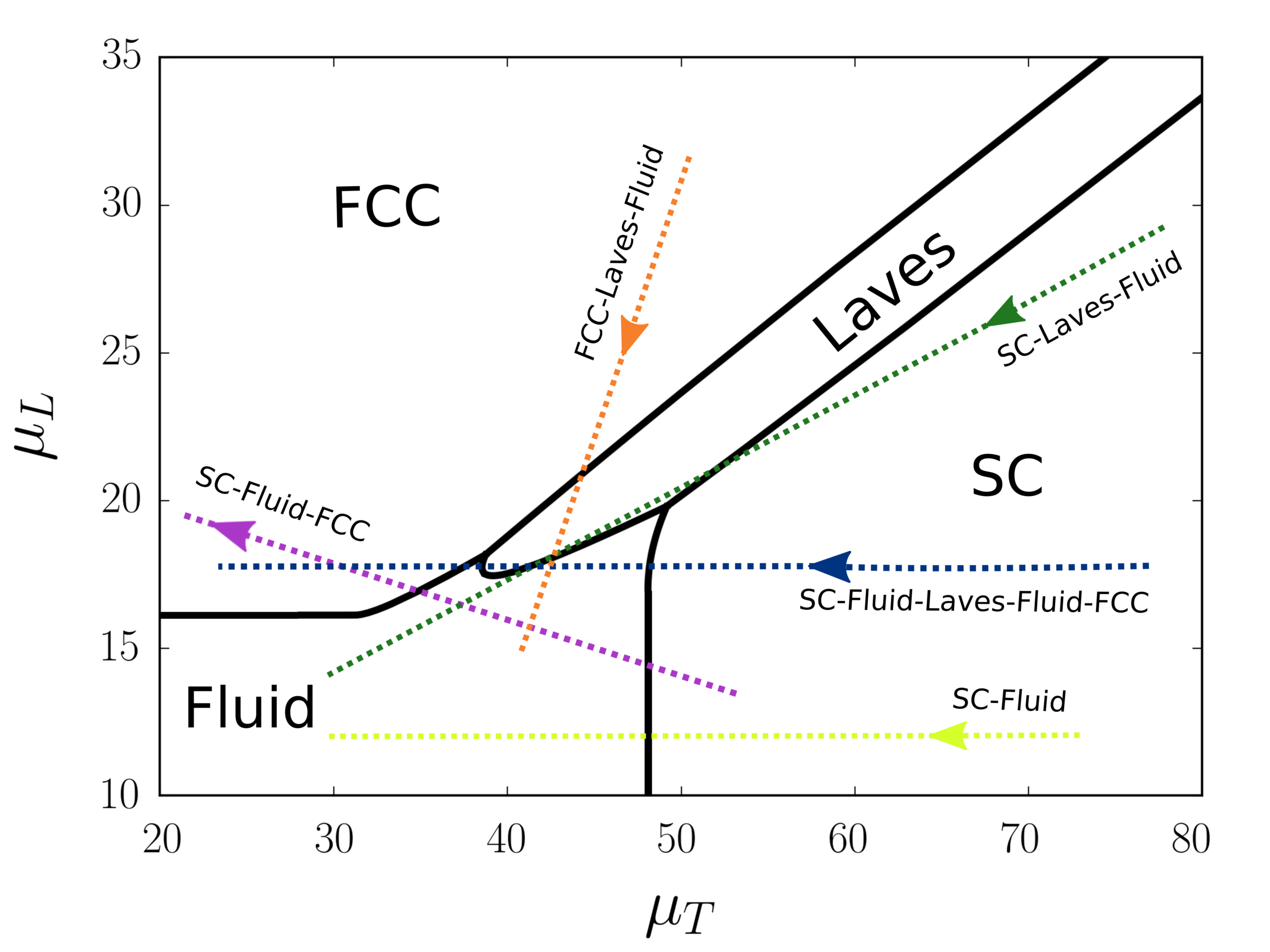}
  \caption{{\small Phase diagram of the binary mixture in the chemical potential of large hard spheres $\mu_L$ -- chemical potential of hard tetramers $\mu_T$ representation. The solid lines represent the bulk binodals, and delimit single-phase regions. The phase transitions of the pure system of spheres and the pure system of tetramers are shown by the horizontal and vertical asymptotic extensions of the respective binodals. The colored dashed lines represent some of the possible phase-stacking sequences in the sediment. The color code is the same as the one used for the stacking diagram. The arrows on the dashed lines indicate the direction from the bottom to the top of the sediment.}}
  \label{fig:mus_mut}
\end{figure}

~We now study the system while sedimenting under a gravitational field. The phase diagram in the chemical potential of the spheres $\mu_L$ -- chemical potential of the tetramers $\mu_T$ representation is shown in Fig.~\ref{fig:mus_mut}, where full black lines represent bulk binodals. At each point on a binodal two phases are in equilibrium with each other. 

This bulk phase diagram is used as an input for our theory, as discussed in Sec.~\ref{ssec:stackmeth}, in order to calculate the stacking diagram. The different regions in a stacking diagram, each of which represent a unique stacking sequence, are delimited by the following features:
 
\textbf{Sedimentation binodal} is the locus of all sedimentation paths tangential to the bulk binodal(s). We have five bulk binodals indicating the various coexistences as shown in Fig.~\ref{fig:mus_mut} thus giving five corresponding sedimentation binodals.

\textbf{Terminal lines} which represent sedimentation paths passing through any point where a binodal terminates. As can be seen from Fig.~\ref{fig:mus_mut}, we have three such terminal points:
\begin{enumerate}
	\item The triple point where the Laves, fluid and FCC phases coexist.
	\item The triple point where the Laves, fluid and SC phases coexist.
        \item The triple point where the Laves, FCC and SC phases coexist. In order to locate this point in the plane of chemical potentials, we obtain the FCC-Laves and SC-Laves binodals from simulations until pressures $\beta P\sigma_L^3=70$, and we extrapolate the last simulated points until the two binodals meet. 
\end{enumerate}

\textbf{Asymptotic terminal lines} appear when the bulk binodal does not terminate at a finite value for one of the chemical potentials, i.e, when the binodal is connected to a phase transition of a one-component system or when, at very high chemical potentials, both pure component crystals approach close packing densities. For example, in our system the fluid-FCC binodal goes asymptotically to the fluid-FCC phase transition of the pure large hard spheres, which is denoted by the $s=0$ line. In addition, the fluid-hard tetramer SC phase continues in the pure tetramer system, which is at $s = -\infty$ (not shown in figure).

\begin{figure}[htb]
  \centering
  \includegraphics[width=0.5\textwidth]{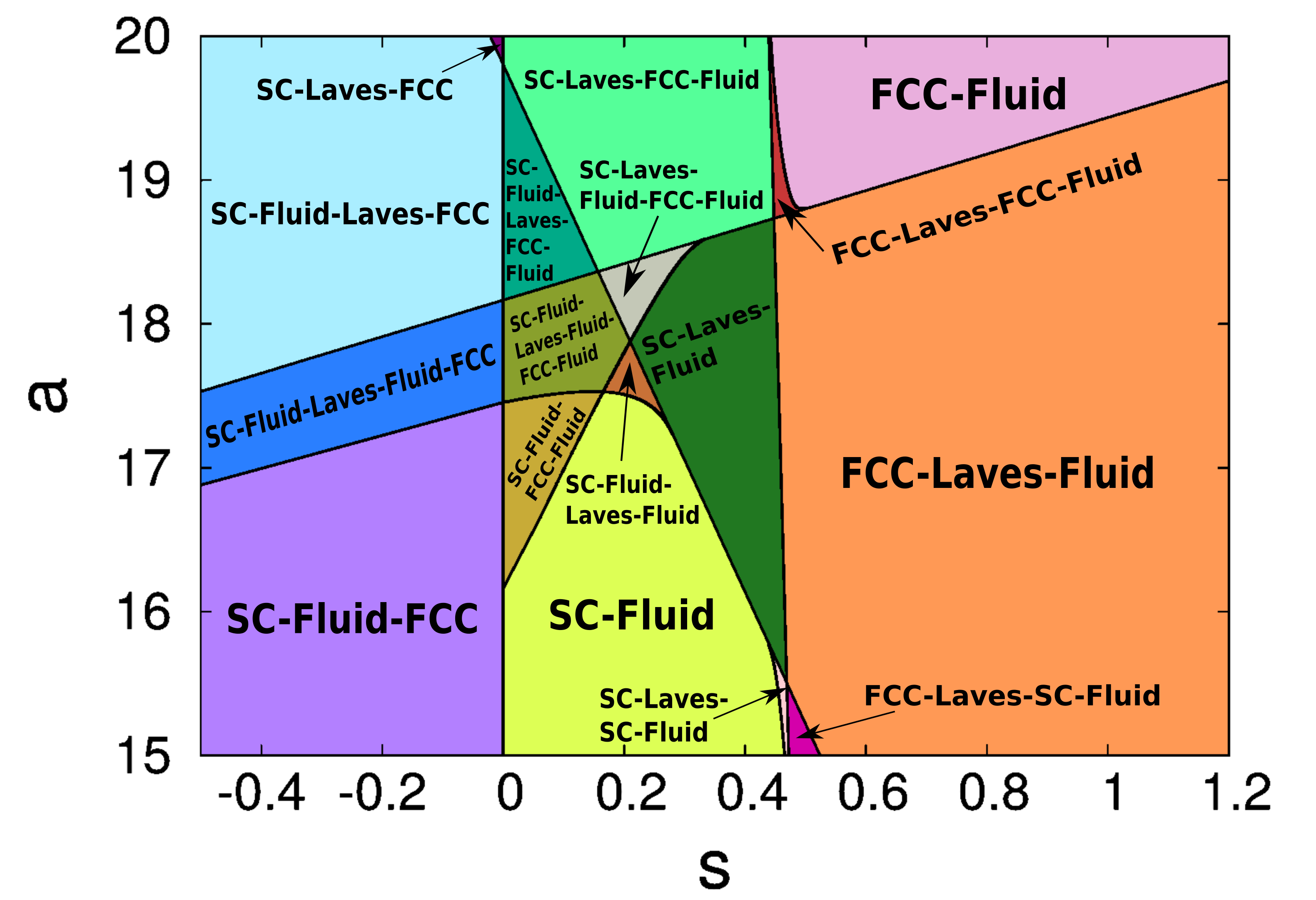}
  \caption{{\small The stacking diagram depicting the sedimentation-diffusion equilibrium for the binary mixture of large hard sphere-hard tetramer. The colored regions represent the different phase-stacking sequences of phases observed in the sedimentation column. Note that the colors of the regions correspond to the colors of the sedimentation paths drawn on the $\mu_T$ -- $\mu_L$ representation in Fig. \ref{fig:mus_mut}. For visualisation purposes we have restricted the axes to the region which contains the largest amount of stacking sequences. Moreover, the $a$ axis was linearly scaled with respect to $s$ by a constant $c=-40$, such that $a = a_{actual} - c\; s$.}}
  \label{fig:a_s}
\end{figure}

With these features in place, we obtain the corresponding stacking diagram of system of large hard spheres and hard tetramers undergoing sedimentation, shown in Fig.~\ref{fig:a_s}, with the assumption that both species sediment slowly enough for LDA to apply. The differently colored regions in Fig.~\ref{fig:a_s} represent the different stacking sequences for this binary mixture. We remind the reader that $s$ equals the ratio of the buoyant masses of the spheres to the tetramers, as from Eq.~\ref{SD3}. A negative $s$ means that one species settles while the other creams up. For the purposes of analysis, in this paper we assume that the tetramers always settle, which means that the buoyant mass of the tetramer species is always positive. Alternatively, if the identity of the settling species is switched, the stacking sequences for the negative $s$ region will simply be reversed. Keeping this in mind, the following observations can be made about the resulting stacking diagram. 
\begin{enumerate}
\item For negative $s$, the single species crystal phase formed at the bottom is always the SC phase of tetramers, as is expected. 
\item For $0 < s \lesssim 0.45$, the pure component crystal phase is the SC of tetramers, which is also expected because the tetramers have a higher buoyant mass than the spheres and therefore sediment faster.
\item  For $s \gtrsim 0.45$, the large spheres form the FCC phase at the bottom of the column. This is counter-intuitive as the spheres have a lower buoyant mass and should sediment less than the tetramers, and it is reminiscent of the ``Brazil-nut effect" in binary granular mixtures under shaking, where the large species rises to the top of the smaller one~\cite{bib:williams-segr.review,bib:rosato-br.nuts}. However, the nature of the two phenomena is different, because the Brazil-nut effect happens under out of equilibrium conditions, while the observed settling behaviour is an equilibrium phenomenon. We also note that Brazil-nut-like effects have been observed -- both theoretically and experimentally -- in colloidal systems, however involving charged binary mixtures.~\cite{bib:loewen-br.nut,bib:vanderlinden-colloid.br.nut}
\item For $s \approx 1$, which means that the buoyant mass of both species are equal, we still observe that the large hard spheres form an FCC crystal at the bottom. This can be understood if we approximate the hard tetramer by a circumscribed sphere with a diameter $\sigma_{\textrm{eff}}=q(1+\sqrt(3/2))\sigma_L\simeq 1.83\sigma_L$. We thus find that the tetramers are larger in size than the spheres, and hence the system minimises its potential energy by having the smaller species at the bottom.
\item  We intriguingly observe some regions with floating crystal phases~\cite{bib:delasheras-float.phases}, where crystaline phases are found on in between the fluid phase, such as SC-Fluid-Laves-Fluid or SC-Fluid-FCC-Fluid. These regions are, however, relatively small.
\end{enumerate}

\section{\label{sec:summary}Conclusions}
We investigated the phase behaviour of a binary mixture of hard spheres and hard tetramers consisting of beads arranged in a tetrahedral fashion. By using MC simulations in the isobaric-isothermal ensemble combined with free-energy calculations and the thermodynamic integration method, we mapped out the bulk phase diagram of the mixture in the pressure $\beta P\sigma_L^3$--composition $x$ representation. We also theoretically determined the sedimentation behaviour of this mixture using the local density approximation.

We found two-phase coexistence regions between the fluid phase and the various crystal structures, as well as two triple points, namely the Fluid+SC+Laves and the Fluid+Laves+FCC triple points. Surprisingly, we find a relatively large coexistence region between the fluid and the Laves phase -- the structural analogue of the \ce{MgCu2} phase, which may be experimentally accessible. In order to facilitate comparison with experimental parameters, we also converted the phase diagram from the pressure $\beta P\sigma_L^3$ -- composition $x$ representation to the packing fraction of tetramers $\eta_T$ -- packing fraction of spheres $\eta_L$ plane.

Assuming the validity of the local density approximation under relevant experimental conditions for our binary system, we also investigated the sedimentation behaviour by calculating the stacking diagram of this mixture. We observed several stacking sequences, some of which were reminiscent of the ``Brazil-nut effect'' in binary granular mixtures, while others intriguingly displayed floating crystal phases.

Our results demonstrate a novel self-assembly route towards a photonic crystal, in which the Diamond and the Pyrochlore structures can be assembled in one crystal -- the \ce{MgCu2} Laves structure -- from a binary mixture of hard spheres and hard tetramers. By selectively burning or dissolving one of the species, either the tetramers or the spheres, the Laves phase can be converted into a diamond lattice or a pyrochlore structure to obtain a photonic crystal with a bandgap in the visible range. We hope that our results will stimulate further experimental and theoretical investigations. In future work, we will address the crystallization kinetics of the proposed self-assembly route, as well as the effect of colloidal epitaxy. 

\section*{Acknowledgements}
This work is part of the research programme of the Foundation for Fundamental Research on Matter (FOM), which is part of the Netherlands Organisation for Scientific Research (NWO). G.A. thanks L. Filion and S. Dussi for fruitful discussions. The authors thank H. Pattabhiraman and V. Prymidis for critically reading the manuscript.

\bibliography{references} 
\end{document}